\documentclass[twocolumn,rmabstract]{theme/bioRxiv}

\usepackage{amsmath,amssymb}
\usepackage{multirow}
\usepackage{array}
\usepackage{booktabs}
\usepackage{enumitem}
\usepackage{cuted}
\usepackage{changepage}
\setlist{nosep}

\makeatletter
\renewcommand\author[2][]{\gdef\@author{#2}}

\newcommand{\abscontent}{}
\newcommand{\setabstract}[1]{\renewcommand{\abscontent}{#1}}

\renewcommand\@maketitle{%
  \null
  \vskip 0.5em%
  \begin{center}%
    {\LARGE\mdseries\@title\par}%
    \vskip 1em%
    {\large\lineskip .5em%
      \begin{tabular}[t]{c}\@author\end{tabular}\par}%
    \vskip 1.5em%
    {\normalsize\@date\par}%
    \vskip 0.5em%
  \end{center}%
  \par\vskip 0.8em%
  \begin{center}{\Large\bfseries Abstract\par}\end{center}%
  \vspace{0.2em}%
  \begin{center}%
  \begin{minipage}{0.85\textwidth}%
    \normalfont\normalsize\noindent\abscontent%
  \end{minipage}%
  \end{center}%
  \par\vskip 0.8em%
}
\makeatother

\renewcommand{\footnoterule}{%
  \kern -3pt
  \hrule width 0.4\columnwidth height 0.4pt
  \kern 2.6pt
}

\usepackage{titlesec}
\titleformat{name=\section,numberless}
  {\large\sffamily\bfseries}{}{0em}{#1}[]
\titleformat{\section}
  {\large\sffamily\bfseries}{}{0em}{Supplementary Note \thesection: #1}[]
\titleformat{\subsection}
  {\small\sffamily\bfseries}{\thesubsection.}{0.5em}{#1}[]
\titleformat{\subsubsection}[runin]
  {\footnotesize\sffamily\bfseries\itshape}{\thesubsubsection.}{0.5em}{#1. }[]

\begin{document}

\leadauthor{Otsubo et al.}
\shorttitle{Partial exploiters sustain cooperation: the hump-shaped strategy
stably coexists with unconditional cooperators}

\title{Partial exploiters sustain cooperation: the hump-shaped strategy stably coexists with unconditional cooperators}

\author{%
Kai Otsubo$^{1,2,\ast,\dagger}$\thanks{$^{\ast}$These authors contributed equally to this work.}\thanks{$^{\dagger}$Correspondence: k.otsubo686@gmail.com} \and
Yuta Kido$^{1,2,\ast}$ \and
Ryutaro Mori$^{3}$
\thanks{Author contributions: K.O., Y.K., and R.M. designed research; K.O., Y.K., and R.M. performed research; Y.K. analyzed data; K.O., Y.K., and R.M. wrote the paper.}\\[0.5em]
\small $^1$Graduate School of Humanities and Sociology, The University of Tokyo, Tokyo 113-0033, Japan\\
\small $^2$Japan Society for the Promotion of Science, Tokyo 102-0083, Japan\\
\small $^3$Computational Group Dynamics Collaboration Unit, RIKEN Center for Brain Science, Saitama 351-0198, Japan%
}

\date{June 25, 2026}

\setabstract{From collective hunting to environmental problems, social dilemmas are pervasive in human societies. Prior research has documented highly heterogeneous behavioral patterns in such settings. However, how this heterogeneity emerges and how it shapes large-scale cooperation remain unclear. Here, we focus on a robustly observed but underexplored pattern: the hump-shaped strategy (Hump). Individuals adopting Hump match others' contributions up to a threshold, only to reduce their own above it. Using agent-based simulations across group sizes and production-function shapes, we find that Hump is individually adaptive, especially in intermediate-sized groups with step-like production functions. Despite its exploitative nature, Hump also elevates population-level cooperation. The underlying mechanism is that Hump can form a stable equilibrium with unconditional cooperators (AllC), which jointly exclude defectors across a broad range of environments. Our findings suggest that underexplored patterns of behavioral heterogeneity---including both Hump and AllC---play a functional role in sustaining large-scale cooperation.}

\maketitle

\section*{Introduction}
\begin{figure*}[!t]
\centering
\includegraphics[width=\textwidth]{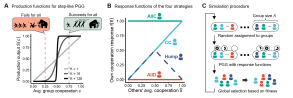}
\caption{\textbf{Overview of the model}
\textbf{(A)} Illustration of the production function for step-like PGG. In addition to a step-like production function ($K=128$), production functions with other values of the steepness parameter $K$ ($K=1, 16$) are shown for reference.
\textbf{(B)} Strategies used in the simulations. The figure shows the response functions of the four strategies---Hump, CC, AllC, and AllD---mapping others' mean cooperation ($\bar{X}$) onto the focal player's cooperation rate.
\textbf{(C)} Simulation Procedure. In each generation, agents are randomly assigned to groups of size $N$ (top row), play the public goods game (middle row; gear icons indicate contributions), which determines each agent's fitness. The next generation is then drawn from the global population pool with probability proportional to fitness (bottom row).}
\label{fig:fig1}
\end{figure*}

\noindent Throughout history, humans have often faced the challenge of producing collective goods. When a group of hunters engages in collective hunting, for example, individuals have an incentive to avoid active participation to reduce personal risk. Yet if everyone behaves this way, the hunt fails (Fig.~\ref{fig:fig1}A). Such conflicts between individual and collective interests have been conceptualized as social dilemmas or collective action problems \cite{Dawes1980-ov, Van_Lange2013-we, Olson1965-lb}. Human behavior in such situations has been studied extensively using game-theoretic paradigms, particularly public goods games (PGGs), in which each member of a group decides how much to contribute to a common pool that benefits all group members. Standard economic theory predicts that self-interested individuals will not contribute to the common pool \cite{Camerer2003-bgt, Ledyard1995-ys}.

Nevertheless, human behavior systematically departs from this prediction: rather than uniformly defecting, many individuals cooperate, often in ways that depend on others' behavior \cite{Ledyard1995-ys, Zelmer2003-oz, Fischbacher2001-lm, Kurzban2005-kg}. By asking participants to specify their intended cooperation levels for each possible level of others' cooperation, prior studies have identified several distinct behavioral types, from unconditional defection and cooperation to various forms of contingent response \cite{Fischbacher2001-lm, Thoni2018-qm}. The rationale behind this heterogeneity has been a long-standing question across biology, economics, and psychology \cite{Grujic2012-se, Bogaert2008-wr,Chaudhuri2011-ux}. One possibility is that this heterogeneity arises because different behavioral strategies are adaptive under different social conditions. 

A natural question, then, is what adaptive value different strategies may have. One well-understood case is conditional cooperation, one of the most prominent and extensively studied patterns in social dilemmas \cite{Chaudhuri2011-ux, Frey2004-zi, Rustagi2010-hw, Kocher2008-hb, Martinsson2013-zs, Herrmann2009-tc, Fischbacher2012-he, Volk2012-vf, Fischbacher2010-tg, Keser2000-ew, Kurokawa2009-qd, Mori2024-wn}. Under this strategy, individuals increase their contributions as others contribute more. Although such behavior is not rational in the one-shot PGGs used in most experimental studies, it can be adaptive in repeated dyadic interactions, allowing individuals to sustain mutual cooperation while avoiding exploitation by defectors \cite{Axelrod1980-so, Axelrod1981-qp, Efferson2024-pk}. Since repeated dyadic interactions are common in everyday life and were likely pervasive in ancestral environments, people may bring this strategy into the laboratory as a heuristic \cite{Li2018-zx, Rand2014-pb, Peysakhovich2016-sp}.

By contrast, no comparable rationalization has been offered for another conditional pattern: the hump-shaped strategy \cite{Fischbacher2001-lm}. Like conditional cooperators, individuals with this strategy increase their cooperation as others' average cooperation rises. Crucially, however, they do so only up to a threshold near 50\%; beyond that point, they instead reduce their cooperation, yielding a hump-shaped response function (Fig.~\ref{fig:fig1}B). Although observed in only about 10\% of participants across samples \cite{Thoni2018-qm}, this pattern has been replicated across diverse settings, ranging from laboratory experiments across multiple countries \cite{Thoni2018-qm, Aimone2013-tp, Kamei2012-pw, Makowsky2014-zb, Fischbacher2001-lm, Martinsson2013-zs, Kocher2008-hb, Herrmann2009-tc} to a field study of forest commons management in rural Ethiopia \cite{Rustagi2010-hw}.

Despite its robust cross-cultural presence, the rationale behind the hump-shaped strategy has never been systematically examined. The logic that rationalizes conditional cooperation does not readily extend to Hump, because those adopting Hump respond to high levels of others' cooperation by reducing their own. This exploitative nature prevents Hump from sustaining mutual cooperation in dyadic settings. Thus, the adaptive basis of the hump-shaped strategy, if any, must lie outside the social environments in which conditional cooperation is adaptive.

Here, we study the adaptive value of the hump-shaped strategy. We propose that it depends on two key features of the environment: the size of the group and the shape of the production function. First, group size may play a key role. In \textit{N}-person public goods games, group size affects the behavioral composition of groups: larger groups are more heterogeneous and more likely to contain a mix of strategies. While those adopting Hump cannot sustain mutual cooperation in dyadic settings, larger and more heterogeneous groups may create conditions under which their response function becomes adaptive.

Second, the steepness of the production function may also matter. It shapes how additional cooperation yields collective goods. At one extreme, a linear function returns a constant amount per unit of cooperation; at the other extreme, a step function produces nothing until a threshold, after which returns are realized in discrete jumps \cite{Kollock1998-xo, Marwell1993-ji}. Previous studies suggest that a step-like structure is a common feature of real-world social dilemmas \cite{Milinski2008-gw, Archetti2012-dm, Archetti2018-dc}. For example, collective hunting also has a step-like structure: if enough members contribute, the hunt succeeds; otherwise, it fails (Fig.~\ref{fig:fig1}A). In such situations, the hump-shaped strategy may be adaptive by avoiding excess contribution while free-riding on collective goods provided by others.

\begin{figure*}[t!]
\centering
\includegraphics[width=\textwidth]{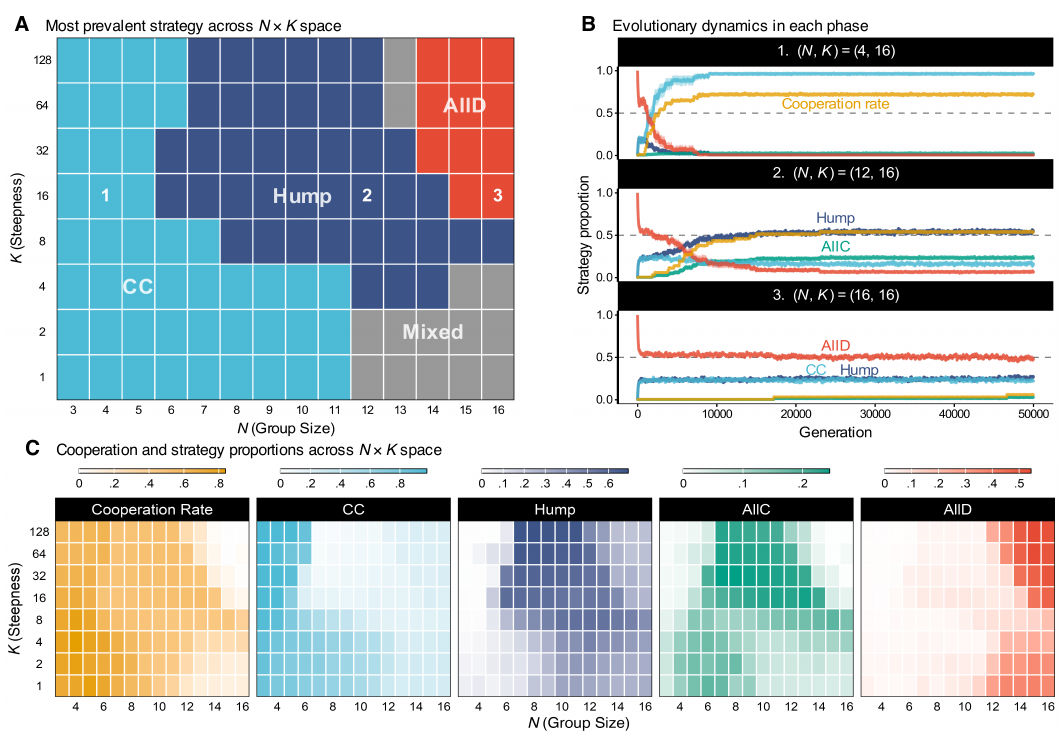}
\caption{\textbf{Three distinct steady-state phases across the $N \times K$ parameter space.}
\textbf{(A)}  Most prevalent strategy phase diagram. Each cell is colored by the strategy with the highest mean steady-state frequency (averaged across 20 runs) if that frequency exceeds 40\%; gray cells indicate no strategy reaches this threshold. Numbered markers (1--3) indicate the representative conditions shown in panel B.
\textbf{(B)} Strategy dynamics at $K = 16$ for three representative group sizes. Lines show cross-run means; shaded bands indicate $\pm$ 1 SE across 20 independent runs. The yellow dashed line shows the population cooperation rate. Top: $N = 4$ (CC-predominant). Middle: $N = 12$ (Hump-predominant). Bottom: $N = 16$ (AllD-predominant).
\textbf{(C)} Cooperation rate and strategy proportions across $N \times K$ space. Five heatmaps display, from left to right: population cooperation rate, and the steady-state frequencies of CC, Hump, AllC, and AllD.}
\label{fig:fig2}
\end{figure*}

In our study, we simulate evolutionary competition among four major strategies identified in past research \cite{Thoni2018-qm}: hump-shaped (Hump), conditional cooperation (CC), unconditional cooperation (AllC), and unconditional defection (AllD) (Fig.~\ref{fig:fig1}B). Agents form \textit{N}-person groups to play PGGs, and the population evolves according to the average fitness of these strategies (Fig.~\ref{fig:fig1}C). Within this simulation, we examine two aspects of Hump. First, by systematically varying group size and production-function steepness, we identify the conditions under which Hump is individually adaptive. Second, by comparing populations that include and exclude Hump, we investigate its impact on population-level cooperation. Finally, we complement these simulation results with a mathematical analysis that reveals the underlying mechanism.

\section*{Results}

\subsection*{Simulation Settings}

We simulate the evolution of four behavioral strategies (AllC, AllD, CC, and Hump; see Fig.~\ref{fig:fig1}B) in an $N$-person public goods game. A population of $250$ groups, each of size $N$, evolves over $50{,}000$ generations. In each generation, group members play the public goods game: AllC contributes at full rate and AllD contributes nothing, while conditional strategies (CC and Hump) adjust their cooperation iteratively in response to others' behavior, starting from an initial belief (see Methods), until convergence. To determine fitness, we use the cooperation rates reached at within-group convergence. We assume that this convergence is negligibly fast relative to evolutionary change \cite{Grujic2012-se, Ohtsuki2009-ir}. Between generations, offspring are drawn with probability proportional to fitness. Each offspring inherits its parent's strategy and initial belief with a small probability of random mutation. The population starts from a uniform AllD population. We crossed group size $N \in \{3, \ldots, 16\}$ and production-function steepness $K \in \{1, 2, 4, \ldots, 128\}$. Below, we report steady-state strategy frequencies averaged over the final 10 generations.

\begin{figure*}[t!]
\centering
\includegraphics[width=\textwidth]{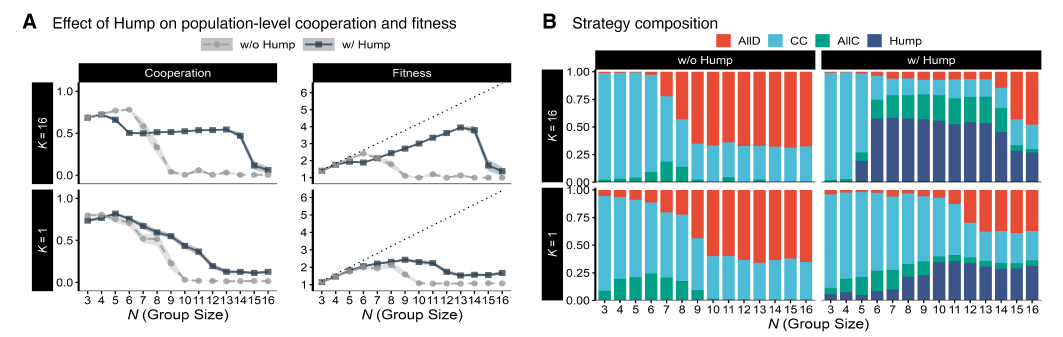}
\caption{\textbf{Effect of Hump on cooperation and social welfare across group sizes.}
\textbf{(A)} Effect of Hump on population-level cooperation and social welfare. Left column: population cooperation rate; right column: mean fitness. Dark squares with solid lines: four-strategy model (with Hump); gray circles with dashed lines: three-strategy model (without Hump). Top row: $K = 16$; bottom row: $K = 1$. The dotted line in the fitness panels indicates the fitness at full cooperation ($\text{MPCR} \times N$).
\textbf{(B)} Strategy composition. Stacked bar charts show steady-state strategy frequencies as a function of $N$. Left column: without Hump; right column: with Hump. Top row: $K = 16$; bottom row: $K = 1$.}
\label{fig:fig3}
\end{figure*}

\subsection*{Intermediate $N$ and high $K$ favor Hump strategy}

Fig.~\ref{fig:fig2}A shows the most prevalent strategy at steady state for each group size $N$ and production-function steepness $K$. We find three distinct phases. First, at small $N$, CC predominates. Second, at large $N$ with high $K$, AllD predominates and cooperation collapses. Finally, between these extremes, Hump predominates at intermediate $N$ with a sufficiently steep production function.

To illustrate the evolutionary dynamics underlying these phases, Fig.~\ref{fig:fig2}B shows strategy trajectories at $K = 16$ for three representative conditions (markers 1--3 in Fig.~\ref{fig:fig2}A). At $N = 4$ (marker 1), CC rapidly spread and predominated ($\approx$ 96\%), achieving high cooperation. At $N = 12$ (marker 2), Hump predominated ($\approx$ 56\%) while AllC persisted as a substantial minority ($\approx$ 23\%), maintaining cooperation at $\approx$ 56\%. At $N = 16$ (marker 3), AllD predominated ($\approx$ 49\%) with Hump ($\approx$ 26\%) and CC ($\approx$ 22\%) as the secondary strategies; AllC collapsed to $\approx$ 2\% and cooperation fell to a low level ($\approx$ 5\%).

Fig.~\ref{fig:fig2}C shows how the cooperation rate and individual strategy frequencies varied across the parameter space. Generally, the cooperation rate declined with increasing $N$. CC was adaptive at small $N$ and low $K$, while Hump became adaptive at intermediate $N$ when the production function was sufficiently steep ($K \geq 8$ and $N \approx 7$--13). Notably, AllC was never the predominant strategy in any condition, yet it retained a 10--25\% minority share precisely in the Hump-predominant region. At large $N$ and high $K$, AllD spread throughout the population, suppressing all other strategies.

\subsection*{Hump presence can promote population-level cooperation}
Having found that Hump can be individually adaptive depending on group size and production function, we now ask how its presence affects cooperation at the population level. One might expect that, because Hump free-rides when others cooperate at high levels, its presence would reduce overall cooperation. To test this, we compare populations that include Hump in their strategy set against those that exclude it, under identical settings.

Instead, adding Hump substantially improved both the population-level cooperation rate and mean fitness across intermediate-to-large group sizes (Fig.~\ref{fig:fig3}A; SI Appendix, Fig.~S1 for the complete sweep across all $K$ conditions). Without Hump, AllD quickly spread as group size increased, driving cooperation to near zero and collapsing mean fitness. With Hump, cooperation was maintained across a much wider range of group sizes, and mean fitness tracked the full-cooperation payoff closely (dotted line in Fig.~\ref{fig:fig3}A, right panel). This effect was most pronounced under steep production functions ($K = 16$), but was also observed under linear production ($K = 1$). When $N$ was small and CC alone could sustain cooperation, the presence of Hump marginally reduced cooperation; however, overall, adding Hump to the strategy set broadened the range of environments in which cooperation was sustained.

Fig.~\ref{fig:fig3}B displays the steady-state strategy composition as stacked bar charts, comparing populations without Hump (left) and with Hump (right) at each $N$ for $K = 16$ (top row) and $K = 1$ (bottom row). The contrast was clearest at $K = 16$. Without Hump, AllD prevailed at $N \geq 9$, whereas with Hump, AllD was suppressed, replaced by a stable Hump--AllC coexistence (Hump $\approx$ 55\%, AllC $\approx$ 24\%) across $N = 8$--13. At $K = 1$, a similar displacement of AllD occurred, though Hump coexisted primarily with CC and AllC. Across all conditions, Hump improved cooperation by stably coexisting with a cooperative strategy (AllC or CC) and jointly excluding AllD.

Taken together, the simulation results show that Hump is individually adaptive at intermediate group sizes under steep production functions (Fig.~\ref{fig:fig2}). Under these conditions, Hump's presence also increased population-level cooperation and mean fitness (Fig.~\ref{fig:fig3}). Common to both patterns was Hump's stable coexistence with AllC. But why does an exploitative strategy coexist with a cooperative one and sustain, rather than undermine, cooperation? To answer this, we now analyze the conditions under which Hump and AllC can stably coexist.

\subsection*{Mathematical Analysis}

\begin{figure*}[t!]
\centering
\includegraphics[width=\textwidth]{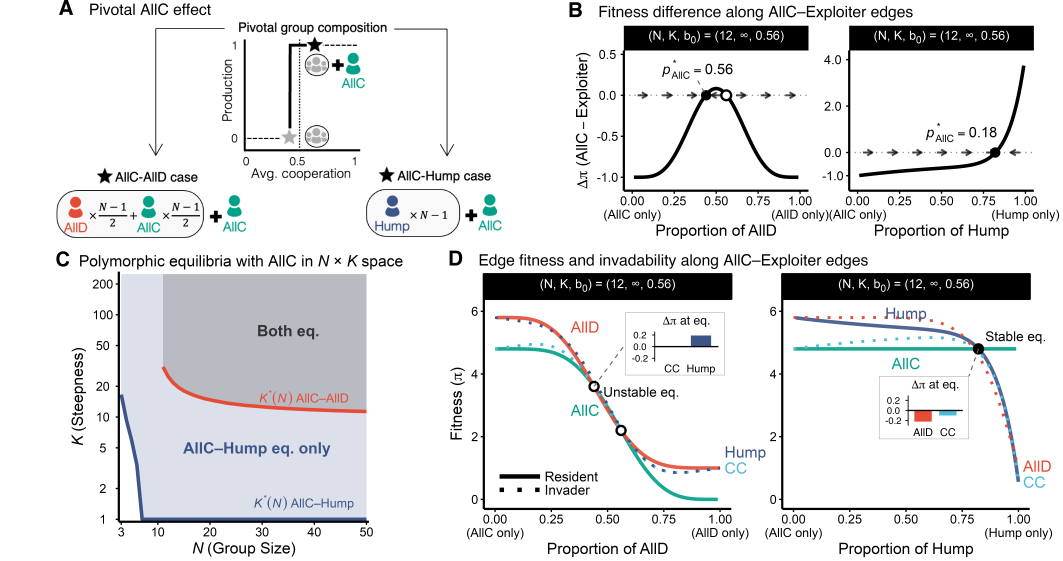}
\caption{\textbf{Structural basis for the AllC--Hump polymorphic equilibrium.}
\textbf{(A)} Pivotal AllC effect. Top: schematic illustrating how a single AllC player can tip the group mean across the production threshold ($x_0 = 0.5$) under step-function production. Bottom: group compositions in which this pivotal effect occurs, shown for the AllC--AllD edge (left) and the AllC--Hump edge (right).
\textbf{(B)} Fitness difference profiles ($\Delta\pi = \pi_\text{AllC} - \pi_\text{Exploiter}$) on two-strategy edges under step-function production ($N = 12$, $K \to \infty$, $b_0 = 0.56$). Left panel: AllC--AllD edge. Right panel: AllC--Hump edge. Filled circles ($\bullet$): stable equilibria. Open circles ($\circ$): unstable equilibria. Arrows: directions of replicator dynamics along each edge.
\textbf{(C)} Equilibrium existence regions in $N \times K$ space. Critical steepness $K^{*}(N)$ for polymorphic equilibrium existence (navy: AllC--Hump; red: AllC--AllD), averaged over $b_0 \in [0, 1]$.
\textbf{(D)} Edge fitness profiles in the full four-strategy system under step-function production ($N = 12$, $K \to \infty$, $b_0 = 0.56$). Left: AllC--AllD edge. Right: AllC--Hump edge. Solid lines: resident strategies. Dotted lines: absent (invader) strategies. Insets: invader relative fitness $\Delta\pi = \pi_\text{invader} - \pi_\text{resident}$ at the marked equilibrium.}
\label{fig:fig4}
\end{figure*}

To understand why AllC--Hump coexistence (i.e., polymorphic equilibrium) is stable, we first analyze populations with only two strategies (\textit{edges}): AllC paired with an exploitative strategy (AllD or Hump). We begin with step-function production ($K \to \infty$) for two reasons: Hump--AllC coexistence was most pronounced at high $K$ in our simulations (Fig.~\ref{fig:fig2}), and this limit is analytically most tractable. Here, a single AllC can produce a discrete jump in production from zero to its maximum (Fig.~\ref{fig:fig4}A). Such jumps occur only in \textit{pivotal compositions}, strategy mixes in which one player's strategy switch tips the group mean across the production threshold. In pivotal compositions, the production gain can exceed the contribution cost, making cooperation individually profitable; elsewhere, defection remains strictly better. A stable coexistence therefore exists only when pivotal compositions are frequent enough that their expected production gain offsets the cooperator's average contribution cost. Prior work has established this mechanism for AllC--AllD coexistence in threshold public goods games \cite{palfrey1984participation,Archetti2011-ev,Kristensen2022-as}. We show below that the same mechanism supports AllC--Hump coexistence under much broader conditions.

\subsubsection*{AllC--Hump coexistence arises across broad conditions}

We first compare the pivotal compositions on the AllC--AllD and AllC--Hump edges. On the AllC--AllD edge, a single AllC player tips the group mean across the production threshold ($x_0 = 0.5$) only when roughly half the group is AllC. The AllC--Hump edge has a fundamentally different structure. An all-Hump group never exceeds the production threshold. Strikingly, however, adding a single AllC player always pushes the group mean across it (see Methods for the proof). The pivotal composition on the AllC--Hump edge is therefore uniquely the all-Hump group (Fig.~\ref{fig:fig4}A).

The contrast between these two pivotal compositions leads to markedly different conditions for polymorphic equilibria. AllC's fitness advantage peaks at the population state where random group sampling most often produces the pivotal composition (i.e., at the mode of the binomial distribution; Fig.~\ref{fig:fig4}B). On the AllC--AllD edge, the pivotal composition (half AllC, half AllD) sits at the center of the binomial distribution; AllC--AllD coexistence therefore requires large groups ($N \geq 11$ at $\mathrm{MPCR} = 0.4$). On the AllC--Hump edge, by contrast, the all-Hump pivotal composition sits at the edge of the distribution, reaching probability close to $1$ in a nearly all-Hump population, so AllC--Hump coexistence exists for all $N \geq 3$ at the same MPCR (see Methods for the derivations). Thus, across the $N \times K$ parameter space, the AllC--Hump equilibrium spans a much broader region (Fig.~\ref{fig:fig4}C).

The second advantage concerns how easily coexistence can emerge from pure-exploiter populations. On the AllC--AllD edge, a single AllC within an all-AllD population is never pivotal. The population therefore remains stuck in full defection (Fig.~\ref{fig:fig4}B, left). On the AllC--Hump edge, by contrast, an AllC player placed in an all-Hump population is always pivotal, so AllC can invade, and coexistence is the unique outcome on the edge (Fig.~\ref{fig:fig4}B, right). 

\subsubsection*{AllC--Hump coexistence is stable against invasion by other strategies}

Thus far, we have analyzed the conditions for AllC--Hump equilibrium between pairs of strategies. We now ask whether this equilibrium remains stable against invasion by the remaining strategies (AllD and CC) in the full four-strategy system. To test this, we compare how each strategy's fitness changes along each two-strategy edge containing AllC (see Methods). We report results at the representative condition ($N = 12$, $K \to \infty$, $b_0 = 0.56$). The SI Appendix reports the analysis over the full $b_0$ range and across $K$ conditions (Fig.~S6), together with a robustness check in which a response slope slightly below one removes the $b_0$ dependence and preserves the AllC--Hump coexistence (Sec.~11, Figs.~S9 and S10).

At this representative condition, AllC--Hump coexistence is uninvadable: both AllD and CC have lower fitness than the residents (Fig.~\ref{fig:fig4}D, right). By contrast, on the AllC--AllD edge, Hump has higher fitness than the residents, so no stable coexistence is possible (Fig.~\ref{fig:fig4}D, left). This pattern holds across the $N \times K$ parameter space: AllC--Hump coexistence spans most of the space and remains stable against invasion, while AllC--AllD coexistence exists only in a narrow corner and is vulnerable to Hump (SI Appendix, Figs.~S5 and S6). The AllC--CC edge has no stable coexistence either; because CC matches AllC's behavior, it is readily invaded by both AllD and Hump (SI Appendix, Fig.~S8). Therefore, across the range of environments studied here, Hump is the only strategy that can form a stable coexistence with AllC.

Our mathematical analysis resolves the paradox raised in the simulation: why an exploitative strategy can promote cooperation at the population level. The answer lies in Hump's response function: adding a single AllC player to an all-Hump group always pushes the group mean across the production threshold. This structural property ensures that Hump and AllC form a stable coexistence across a broad region of the $N \times K$ parameter space, stable against invasion by the other strategies. This coexistence excludes AllD and sustains cooperation. This is the mechanism behind Hump's emergent role as a cooperation-sustaining exploiter in the simulations.

\section*{Discussion}

The literature has documented substantial heterogeneity in behavioral patterns in social dilemmas. Amid such heterogeneity, a subset of individuals adopts a puzzling behavioral pattern: the hump-shaped strategy (Hump) \cite{Fischbacher2001-lm}. Despite consistent observation across diverse cultural contexts  \cite{Aimone2013-tp, Kamei2012-pw, Makowsky2014-zb, Fischbacher2001-lm, Martinsson2013-zs, Kocher2008-hb, Herrmann2009-tc, Rustagi2010-hw, Thoni2018-qm}, it has typically been dismissed as noise or subsumed under other categories \cite{Thoni2018-qm, Volk2012-vf, Thoni2012-ek, Dariel2014-jn, Fosgaard2014-ns, Cubitt2017-lo, Gachter2017-jh, Fallucchi2019-dq} and its adaptive basis has remained poorly understood. To fill this gap, we examined the adaptive value of Hump and its consequences for population-level cooperative dynamics across a broad range of group sizes and production-function steepness (linear to step-like).

We find that Hump becomes prevalent in intermediate-sized groups under steep production functions, where it coexists with AllC while jointly excluding AllD. Through this coexistence, Hump enhances both cooperation and fitness at the population level. Mathematical analysis reveals that this pattern rests on the unique ability of Hump to form a stable polymorphic equilibrium with AllC across a broad range of environments. The key mechanism is that even a single AllC player in an otherwise all-Hump group pushes group cooperation above the production threshold, giving AllC a pivotal advantage that sustains its presence. Although Hump itself can be seen as an exploitative strategy, it plays a functional role in sustaining cooperation within heterogeneous populations.

\subsection*{Rethinking cooperative dynamics with heterogeneous individuals}

Our findings highlight the critical role of interactions among heterogeneous behavioral strategies in enabling successful collective action. Although past research has documented heterogeneity in behavioral patterns in social dilemmas \cite{Ledyard1995-ys, Zelmer2003-oz, Fischbacher2001-lm, Kurzban2005-kg}, less attention has been paid to how this heterogeneity contributes to cooperative dynamics. Our findings imply that large-scale cooperation in human societies may be sustained, at least in part, by interactions among individuals with diverse behavioral strategies. This perspective aligns with previous research suggesting that observed heterogeneity may reflect the stable coexistence of functionally distinct types at equilibrium \cite{Kurzban2005-kg, Grujic2012-se}. Furthermore, we show that this mechanism is particularly important in more ecologically relevant settings, especially those involving larger groups and nonlinear production functions \cite{Milinski2008-gw, Archetti2012-dm, Archetti2018-dc, Levin2014-pg}. Given that Hump-like strategies have been shown to destabilize cooperation in dyadic, linear settings \cite{Efferson2024-pk}, moving beyond such settings can give rise to patterns of cooperation not predicted by conventional models. 

A notable finding of this study is that an exploitative strategy (Hump) can coexist with a fully cooperative strategy (AllC). In our simulations, AllC persists as a stable minority in the Hump-predominant region, and this persistence sustains population-level cooperation. This finding sheds light on the role of AllC, a strategy generally considered non-adaptive because it is readily exploited by defectors \cite{Nowak2006-wh}. The relationship between AllC and Hump is reminiscent of host--parasite coevolution \cite{Anderson1982-fa}: the exploited partner can persist despite exploitation, thereby maintaining the conditions under which exploiters can exist. Thus, the AllC--Hump coexistence may reflect a broader feature of cooperative systems in biology. 

These results may also have practical implications for promoting real-world collective action \cite{Kraft-Todd2015-ey}. Many real-world public goods have a threshold-like rather than linear structure \cite{Milinski2008-gw}. In such settings, interventions need not eliminate all free-riding. Instead, tolerating partial free-riding may yield better collective outcomes than attempting to enforce uniformly high cooperation in some contexts. Field evidence is consistent with this view. In a long-term study of threshold-structured group lending, outright free-riding was rare, and in most groups some members partially free-rode while others contributed just enough to carry the group over its threshold \cite{Sabin2026-pd}. Institutions that sustain such a coexistence of heterogeneous contributors may therefore accomplish more than those that maximize every member's contribution.

\subsection*{Limitations}
Several limitations warrant consideration. We assumed a limited set of behavioral strategies, Hump, CC, AllC, and AllD, based on a well-established typology in the literature \cite{Fischbacher2001-lm, Thoni2018-qm}. However, some recent research proposes alternative ways to categorize strategy types, for example, using machine-learning-based methods \cite{Fallucchi2019-dq, Wang2025-ah}. Moreover, classified participants often deviate from the prototypical response pattern of their assigned type \cite{Fischbacher2001-lm}, casting doubt on the assumption of discrete types itself. Future research should examine whether our results hold under continuous strategy formulations.

Moreover, our results partially depend on the alignment between Hump's cooperation threshold ($\theta$) and the inflection point of the production function ($x_0$). Supplementary analyses show that when Hump starts to reduce its cooperation before the threshold is reached ($\theta < x_0$), the strategy becomes individually non-adaptive and can undermine collective cooperation. When $\theta \geq x_0$, however, the AllC--Hump coexistence holds regardless of $x_0$ (SI Appendix, Sec.~12 and Figs.~S11--S15). The adaptive value of Hump thus rests on a clear necessary condition: its behavioral rule must be appropriately tuned to the threshold structure of the environment. Whether individuals adopting Hump actually calibrate their thresholds to the production structure they face remains an open empirical question. 

\subsection*{Conclusion}
This study investigates the adaptive value of the puzzling behavioral pattern, Hump. Through agent-based simulations, we show that under specific environmental conditions---intermediate-sized groups with step-like production functions---Hump can be individually adaptive and help sustain cooperation at the population level. These results likely reflect the unique ability of Hump to form a stable coexistence with AllC and resist AllD invasion in a broad range of environments. Our results suggest that observed heterogeneity in behavior, including Hump and AllC, may be a key to successful cooperation.

\section*{Materials and Methods}
\subsection*{Public Goods Game}
We model a repeated $N$-person public goods game. In each interaction, each of $N$ players simultaneously chooses a cooperation rate $c_i \in [0, 1]$, representing the fraction of their endowment contributed to a public pool. Contributions are transformed through a production function and the resulting public good is distributed equally among all group members.

\subsection*{Strategies}
Four behavioral strategies, each defined by a response function $f(x)$, govern how players choose their cooperation rate conditional on the mean cooperation rate of their $N-1$ co-players $x$. In the simulation, $x$ corresponds to the observed mean cooperation rate of co-players from the previous round; in the initial round, players use an initial belief $b_0$ (see Agent-Based Simulation below). AllC ($C$ in equations) always fully cooperates: $f(x) = 1$. AllD ($D$) always defects: $f(x) = 0$. Conditional Cooperation (CC) matches others' mean cooperation rate: $f(x) = x$. The Hump-shaped strategy (Hump; $H$ in equations) is a piecewise-linear response function with threshold $\theta = 0.5$:

\begin{equation}
f_H(x) =
\begin{cases}
x & \text{if } x \le 0.5 \\
1 - x & \text{if } x > 0.5
\end{cases}
\label{eq:fH}
\end{equation}
yielding a maximum cooperation of 0.5 at $x = 0.5$ (Fig.~\ref{fig:fig1}B).

\subsection*{Production Function}
The production function $S(\bar{c})$ maps the group mean cooperation rate $\bar{c}$ to a return factor applied to contributions. We use a sigmoid (logistic) function, linearly rescaled so that $S(0) = 0$ and $S(1) = 1$:

\begin{equation*}
S_{\text{raw}}(\bar{c}; K, x_0) = \frac{1}{1 + \exp\left(-K(\bar{c} - x_0))\right)}
\end{equation*}
Two parameters control the shape (Fig.~\ref{fig:fig1}): the steepness $K$, which controls whether the function is near-linear or step-like, and the inflection point $x_0$, fixed at 0.5 in the main analysis (sensitivity analyses with $x_0 = 0.4$ and $0.6$ are reported in SI Appendix, Sec.~12).

\subsection*{Payoff Function}
Each player $i$ with an initial endowment of $1$ and in a group with mean cooperation rate $\bar{c}$ receives:

\begin{equation*}
\pi_i = 1 + \text{MPCR} \times N \times S(\bar{c}; K, x_0) - c_i
\end{equation*}
where MPCR $= 0.4$ is the marginal per capita return and $N$ is the group size. Since MPCR $< 1$ and MPCR $\times N > 1$ for all $N \geq 3$, individual incentives favor defection, yet mutual cooperation yields higher payoffs for all. The full-cooperation payoff is $\pi^* = \text{MPCR} \times N$ (dotted line in Fig.~\ref{fig:fig3}A).

\subsection*{Agent-Based Simulation}
We simulate a population of $M = 250$ groups of size $N$, evolving over $T = 50{,}000$ generations. The parameter sweep spans group size $N \in \{3, \ldots, 16\}$ and steepness $K \in \{1, 2, 4, \ldots, 128\}$ (20 seeds per condition). To assess Hump's impact, we compare a full four-strategy model (AllC, AllD, CC, Hump) against a reduced three-strategy model excluding Hump. The most prevalent strategy in each condition is defined as the strategy with the highest mean steady-state frequency, provided it exceeds 40\%; conditions where no strategy reaches this threshold are classified as mixed.

Within each generation, agents interact repeatedly in their assigned groups. Each agent $i$ carries a belief $b_i^{(t)} \in [0, 1]$ about the mean cooperation rate of co-players, updated each round. Each round $t \geq 1$, agent $i$ cooperates at rate $c_i^{(t)} = f_s(b_i^{(t-1)})$, where $b_i^{(0)}$ is the initial belief inherited from the parent generation with noise (see below) and $b_i^{(t)} = \bar{c}_{-i}^{(t)}$ for $t \geq 1$, with $\bar{c}_{-i}^{(t)}$ denoting the mean cooperation rate of $i$'s co-players in round $t$. Rounds iterate until cooperation rates converge (SI Appendix, Sec.~2 for the convergence criterion). The converged payoffs serve as fitness for between-generation selection.

Between generations, three processes occur. First, fitness-proportional selection draws the next generation from the global population pool, ignoring group boundaries; each offspring inherits its parent's strategy and initial belief, with small Gaussian noise added to the belief ($\sigma = 0.01$, clamped to $[0, 1]$). Second, each agent independently mutates to a uniformly random strategy with probability $\mu = 0.01$. Third, individuals are randomly reassigned to new groups of size $N$.

All agents are initialized with the AllD strategy and initial belief $b_i^{(0)} = 0$; cooperation can arise only through mutation and selection.

\subsection*{Mathematical Framework}
To complement the simulation, we analyze deterministic evolutionary dynamics using the finite-group replicator equation \cite{Archetti2011-ev}, in which an infinite population interacts in finite groups of size $N$.

\subsubsection*{Replicator dynamics}
The replicator equation takes the standard form:

\begin{equation*}
\frac{dp_s}{dt} = p_s \left( \pi_s(\mathbf{p}) - \bar{\pi}(\mathbf{p}) \right)
\end{equation*}
where $p_s$ is the frequency of strategy $s$, $\pi_s(\mathbf{p})$ is the expected payoff of strategy $s$ given population state $\mathbf{p} = (p_C, p_D, p_{CC}, p_H)$, and $\bar{\pi}(\mathbf{p}) = \sum_s p_s \pi_s(\mathbf{p})$ is the population mean payoff.

The expected payoff is computed from the focal-player perspective: a focal individual of type $s$ is grouped with $N-1$ co-players drawn by multinomial sampling from $\mathbf{p}$. Within-group dynamics are iterated from initial belief $b_0$ until convergence, and the focal player's payoff is averaged over all possible co-player compositions (SI Appendix, Sec.~3 for the full formulation). Unlike in the simulation, where beliefs evolve endogenously, the mathematical analysis fixes $b_0$ exogenously and sweeps over $b_0 \in (0, 1)$. Alternatively, reducing the conditional response slope to $1 - \epsilon$ ($\epsilon \to 0^{+}$) eliminates the $b_0$ dependence altogether (SI Appendix, Sec.~11).

\subsubsection*{Polymorphic equilibrium conditions on AllC edges}

We derive equilibrium conditions in the step-function limit ($K \to \infty$), where the production function reduces to a threshold at $x_0 = 0.5$ and the equilibrium conditions admit closed-form expressions. On each two-strategy edge, AllC is paired with an exploiter strategy $A$ (AllD or Hump), whose frequency we denote by $\alpha$. The group-composition-level fitness difference between $A$ and AllC is
\begin{equation}
\delta_k = \Delta\mathrm{cost}_k + \text{MPCR} \times N \times \Delta S_k
\label{eq:delta_k}
\end{equation}
where $k$ counts co-players of type $A$, $\Delta\mathrm{cost}_k$ is the cost saving from playing $A$ over AllC, and $\Delta S_k$ is the resulting change in production when the focal player switches from AllC to $A$. So $\delta_k > 0$ means $A$ is favored at that composition. The population-level fitness difference averages Eq.~\ref{eq:delta_k} over binomial co-player draws:
\begin{equation}
\delta(\alpha) = \sum_{k=0}^{N-1} \binom{N-1}{k} \alpha^k (1-\alpha)^{N-1-k} \, \delta_k
\label{eq:delta_alpha}
\end{equation}

$\Delta S_k$ in Eq.~\ref{eq:delta_k} is nonzero only at \emph{pivotal compositions}, group compositions in which switching the focal player's strategy moves the group mean across the production threshold $x_0 = 0.5$ (Fig.~\ref{fig:fig4}A). At all other compositions, $\Delta S_k = 0$ and $\delta_k = \Delta\mathrm{cost}_k > 0$: $A$ (exploiter) is favored by its lower cooperation cost. Thus, in Eq.~\ref{eq:delta_alpha}, the only negative contributions come from pivotal compositions, where the production jump (second term of Eq.~\ref{eq:delta_k}) favors AllC. Since $\delta(0) > 0$, whether an interior equilibrium exists reduces to whether the binomial weight $\binom{N-1}{k^*} \alpha^{k^*} (1-\alpha)^{N-1-k^*}$ assigned to the pivotal composition $k^*$ in Eq.~\ref{eq:delta_alpha} is large enough to drive $\delta(\alpha)$ below zero.

\paragraph*{AllC--AllD edge ($A$ = AllD)}
Both strategies are unconditional, so switching the focal player from AllD to AllC raises the group mean cooperation rate by exactly $1/N$. This switch is pivotal when approximately half the co-players are AllD ($k^* = \lfloor N/2 \rfloor$), a composition most probable near $\alpha \approx 0.5$. Because this pivotal composition sits at the center of the binomial distribution, its probability scales as $O(1/\sqrt{N})$. A polymorphic equilibrium therefore requires sufficiently large $N$; at $\mathrm{MPCR} = 0.4$, the threshold is $N \geq 11$ (SI Appendix, Sec.~4 for the closed-form derivation). When this equilibrium exists, it coexists with a stable all-AllD state, making the edge bistable.

\paragraph*{AllC--Hump edge ($A$ = Hump)}
Unlike AllD, Hump is a conditional strategy whose converged cooperation rate depends on the group composition through within-group dynamics. Consider a group containing $h$ Hump players and $N - h$ AllC players. When at least one AllC is present ($h < N$), the within-group dynamics must converge to a fixed point of Hump's response function. In the low-cooperation regime of Eq.~\ref{eq:fH} ($f_H(x) = x$, applicable when co-players' mean cooperation $\bar{x}_{-i} \leq 0.5$), the fixed-point equation yields $c_H = 1$, contradicting $\bar{x}_{-i} \leq 0.5$. The system therefore converges to the high-cooperation regime of Eq.~\ref{eq:fH} ($f_H(x) = 1 - x$), where the fixed-point equation
$c_H = 1 - \frac{(N - h) + (h - 1)\,c_H}{N - 1}$
yields the converged cooperation rate of Hump:
\begin{equation}
c_H(h) = \frac{h - 1}{N - 2 + h}
\label{eq:cH}
\end{equation}
Substituting Eq.~\ref{eq:cH} into the group mean gives $\bar{c}(h) = \frac{(N^2 - 2N + h)}{N(N - 2 + h)}$, which satisfies $\bar{c}(h) > 0.5$ for all $h < N$: any group containing at least one AllC player always produces a group mean above the production threshold.

For the all-Hump group ($h = N$), the converged cooperation level depends on the initial belief: $c_H = \min(b_0, 1 - b_0)$, giving $\bar{c}(N) = \min(b_0, 1 - b_0) \leq 0.5$. Since $\bar{c}(h) > 0.5$ for all $h < N$ and $\bar{c}(N) \leq 0.5$, the pivotal composition is uniquely the all-Hump group (i.e., $k = N - 1$ Hump co-players for an AllC focal): adding a single AllC player always pushes the group mean above the production threshold (Fig.~\ref{fig:fig4}A).

In a nearly all-Hump population ($\alpha \to 1$), this pivotal composition becomes highly probable: its binomial weight $\alpha^{N-1}$ in Eq.~\ref{eq:delta_alpha} remains $O(1)$. Marginalizing over $b_0$, a polymorphic equilibrium exists whenever $\mathrm{MPCR} \times N > 0.75$, which holds for all $N \geq 3$ at $\mathrm{MPCR} = 0.4$ (SI Appendix, Sec.~4 for the full derivation of the $b_0$-marginalized condition; Fig.~S2 for the phase diagram in $N \times \mathrm{MPCR}$). This equilibrium is globally stable on the edge: both the all-Hump and all-AllC states are unstable, so the interior equilibrium is the unique attractor.

\paragraph*{Extension to finite $K$}
The conditions derived above hold exactly in the $K \to \infty$ limit. As $K$ increases, the sigmoid production function approaches a step function and the equilibrium conditions converge to these analytical values. We determine the critical steepness $K^*(N)$, the minimum $K$ at which a polymorphic equilibrium first appears on each edge, by numerical search over $K$ at each $N$ (Fig.~\ref{fig:fig4}C).

\subsubsection*{Invasion stability analysis}
To assess whether these two-strategy equilibria survive in the full four-strategy system, we compute edge fitness profiles: for each edge, we evaluate the expected payoff of all four strategies as a function of the resident composition (Fig.~\ref{fig:fig4}D; SI Appendix, Sec.~5 for computational details). An absent strategy $s$ can invade when its fitness exceeds that of the residents ($g_s \equiv \pi_s - \pi_{\mathrm{res}} > 0$).

\subsection*{Supporting Information}
The Supporting Information Appendix is available at \url{https://osf.io/jb58v/files/gmksb}.

\subsection*{Data, Materials, and Software Availability}
All code used to generate the results and figures of this manuscript will be made publicly available upon publication.

\subsection*{Acknowledgments}
We thank Yukari Jessica Tham for her valuable feedback on this work. We thank Hidezo Suganuma for his helpful comments on earlier versions of this manuscript.

\subsection*{Competing Interest Statement}
The authors declare no competing interest associated with this manuscript.

\section*{References}
\bibliography{main_references}

\end{document}